\documentclass[structabstract]{aa}  

\usepackage[dvips]{graphicx}
\usepackage{txfonts}
\usepackage{natbib}

\begin{document}

\title{Gravitational wave background from population III binaries}

  \author{I. Kowalska
         \inst{1}
         \and
         T. Bulik\inst{1}
         \and
         K. Belczynski\inst{1,2}
         }

  \institute{Astronomical Observatory, University of Warsaw, Al Ujazdowskie 4,
00-478 Warsaw, Poland
\and Dept. of Physics and Astronomy, University of Texas,  Brownsville, TX 78520, USA
}      

  \date{Received  ; accepted  }

\abstract
  {Current star formation models imply that the binary 
fraction of  population III stars is non-zero. 
The evolution of these binaries must have led to the formation of
compact object binaries.
 }
  {We estimate the gravitational wave background 
originating in these binaries and discuss its observability .}
  {The properties of the population III binaries are 
investigated using a binary population synthesis code.
We numerically model the background and take into account 
the evolution of eccentric binaries.}
  {The gravitational wave background from population III 
binaries dominates the spectrum below 100 Hz. If the binary fraction is larger 
than $10^{-2}$, the background will be detectable by Einstein Telescope (ET), Laser Interferometer Space Antenna (LISA) and 
DECi-Hertz Interferometer Gravitational wave Observatory (DECIGO).}
  {The gravitational wave background from population III binaries 
will dominate the spectrum below 100 Hz. The instruments LISA, ET and DECIGO
should either see it easily or, in the case of non-detection, 
provide very strong constraints on the properties of the population III stars.
}


  \maketitle

\section{Introduction}
Coalescing compact object binaries are among the best candidates for
sources of gravitational waves. They include the double neutron star binaries (DNS), the mixed, black hole neutron star binaries (BHNS) and binary black holes (BBH).
A gravitational wave signal may be detected as single events from nearby
sources, but the signals of more distant objects will overlap contributing to the gravitational wave background (GWB).
The GWB from coalescing compact object binaries has been
investigated by numerous authors  \citep[e.g.,][]{2000ApJ...528L..65T,2001ApJ...552..464B,2011PhRvD..84h4004R,2011ApJ...739...86Z,2011arXiv1106.5555O,2011RAA....11..369R,2011PhRvD..84l4037M}.
Existing gravitational wave detectors Laser Interferometer Gravitational Wave Observatory (LIGO)
and Virgo Gravitational Wave Detector have
achieved their design sensitivity (\cite{2011CQGra..28k4002A}, \cite{2008JPhCS.120c2003K}), and are currently
undergoing transitions to the advanced phase with
a sensitivity increase of a factor 30. Additionally, there are
plans to construct other large-scale detectors
such as Large-scale Cryogenic Gravitational wave Telescope (LCGT) \citep{2010CQGra..27h4004K}. In the more distant future,
third generation detectors such as the Einstein Telescope (ET) \citep{2010arXiv1003.1386V} should
be constructed. Theses instrumental developments
ensure that the investigation of the stochastic backgrounds of gravitational waves
is very important. At lower frequencies such as designed band
of ET ($\sim$ 1Hz), we expect large populations of distant object to be observed. The biggest contribution
in that frequency window will come from massive BBH.
There are no direct observations of this kind of objects, hence there are many uncertainties in the stellar evolution models of the formation of BBH.
Because of the high redsift range of future detectors, there is a possibility that among observed
objects there could be the remnants of the oldest stars in the Universe. Population III stars should be
metal free, thus they should produce black holes of masses reaching hundreds of solar masses.
If our understanding of their evolution is correct, there should be a significant gravitational background from remnants of population III stars.

In this paper, we investigate the possible contribution
to the GWB of population III star binaries. Population III
stars are the first stars in the Universe that formed out of metal-free
material. As such, they had very different properties
from the currently forming stars. The evolution and properties
of single population III stars were investigated by \citet{2001ApJ...550..890B} and
\citet{2008IAUS..255...24T}, who found that single
population III stars are stable up to a few hundred of solar masses.
Moreover, they evolve with little mass loss. The evolution of this massive star will end up either in a pair instability supernova leaving no remnant,
or forming a black hole through a direct collapse.
While no population III stars are known, their properties have been
investigated by numerical simulations of the collapse of metal-free
clouds of gas. These simulations have shown that the initial mass function
of a population III star is skewed towards higher masses than those of either population I or II stars 
\citep{2002Sci...295...93A,2002ApJ...564...23B}. The existence of binaries among
population III stars has been uncertain, although simulations of
the collapse of rotating clouds of gas have found evidence of bar instabilities
and the formation of two protostars \citep{2004ApJ...615L..65S,2008ApJ...677..813M}. Since all known stellar populations 
contain a significant fraction
of binaries, it seems justified to consider the properties of population III binaries.
The evolution of population III binaries was
investigated by \cite{2004ApJ...608L..45B} and \cite{2006A&A...459.1001K}.
They found that the coalescence of BBH binaries originating in population III
stars may be a significant source of gravitational waves detectable by the current and future
gravitational wave interferometers.
In section 2, we analyze the gravitational wave spectrum of eccentric binaries,
and section 3 is devoted to the analysis of properties of population III
binaries. Section 4 contains an outline of the estimate of the
GWB. We present the results in section 5, and in section 6 we discuss them.

\section{Population of stars - models}
The evolution of population III  metal-free stars was
first examined in detail about ten years ago by
\cite{2001NuPhA.688..197H}, \cite{2001ApJ...550..890B}, \cite{2001A&A...371..152M}, \cite{2003A&A...399..617M}, and \cite{2002A&A...382...28S}.
These stars were found to be stable with masses of up
to 500  $\textrm{M}_{\odot}$. Population III star evolve very quickly experiencing nearly no mass loss.
Depending on their initial mass, they may form neutrons stars,
black holes, or leave no remnant at all, because of total star disruption in pair instability supernovae \citep{2002ApJ...567..532H}.
The mass spectrum of population III stars is still a matter of debate.
On the one hand the simulations of metal-free cloud collapse have clearly illustrated the possibility
of forming high mass objects. Searches for the low mass metal-free
stars that are predicted to persist in the Universe have brought no success, which indicates
that the low mass cutoff of the population III initial mass function
was significantly higher than for the current population of stars.
An additional question is whether population III stars
were also formed as binaries. Every known stellar population
contains a significant fraction of binaries, so one should
also expect this to be the case for population III stars.
Numerical simulations of the collapse of rotating
metal-free clouds indicate the appearance of a bar instability
leading to the formation of two protostars in a binary \citep{2004ApJ...615L..65S}.
In recent simulations \citep{2009Sci...325..601T,2010MNRAS.403...45S,2011ApJ...737...75G,2011Sci...331.1040C,2012MNRAS.tmp.2251S}, a similar
result was shown. A massive protostellar cloud develops
spiral structure, instabilities, and leads to the formation of several
multiple systems. Thus, we conclude that there is ample justification to investigate population III binary evolution and its consequences.

We model the evolution of population III binaries using a code
described in \cite{2004ApJ...608L..45B}.  The code uses the evolutionary sequences
based on \cite{2001NuPhA.688..197H}, \cite{2001ApJ...550..890B}, and \cite{2001A&A...371..152M}. 
It includes a detailed treatment of the
calculation of stellar remnants.
The binary evolution calculation involves the analysis of the stability of mass
transfers. The common envelope mass transfers end in a merger if the donor is a
main sequence star, and for giant stars we use the standard
formalism of \cite{1984ApJ...277..355W}.

We assumed that the initial mass function of the primaries
stretches between 10 $\textrm{M}_{\odot}$ and 500 $\textrm{M}_{\odot}$ with the exponent of -2.35.
The mass ratio is drawn from a flat distribution.
The initial orbital separation  distribution is flat
as a logarithm of $a$. The range of initial orbital separations is bounded from below by the requirement that
the stars are not in contact at the zero age main sequence (ZAMS) - $a_{min}=1.3(R_1+R_2)$.
The upper bound is more arbitrary. We chose $a_{max}=10^6 R_{\odot}$.
While \citet{2006ApJ...642..382B} argue that the
upper limit to the initial separation is as large as $5\times 10^7 R_{\odot}$,
we wished to include all systems that interact.
The binaries with initial separations close to or above our upper limit will not interact and if they survive the evolution and form compact object binaries,
they will only contribute to the GW background spectrum below $10^{-11}$Hz.
The uncertainty in the number of binaries due to a variation in the upper cutoff
$a_{max}$ is small because of the logarithmic distribution of initial orbital separations.
We calculated the evolution
of binaries and note that the cases when a compact object binary was formed.
We then traced its evolution due to the emission
of gravitational waves.

\subsection{Properties of compact objects}
\begin{figure}
\label{initial}
\includegraphics[width=\columnwidth]{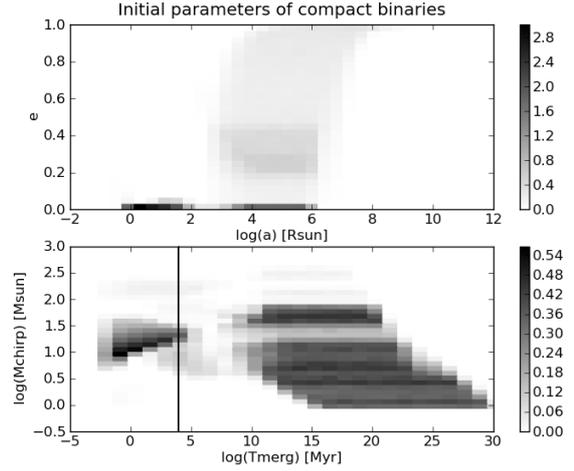}
\caption{The initial properties of the population III compact object binaries.
The top panel shows the distribution of the orbital properties, i.e.
the density in the plane spanned by the semi-major axis and the eccentricity
of the orbit. The lower panel shows the distribution of the chirp masses
and the merger times of the binaries. Vertical line represents the Hubble time. Numbers represent the percentage of all binaries in
the simulation.
}
\end{figure}

We tracked the evolution of $10^6$ binary stars of zero metallicity using the population synthesis code {\tt StarTrack}.
As a result, we obtained 462496 compact objects, which we treat as a realistic sample of population III remnants.
More than 60\% of them are BBH systems, where both components have masses greater than $2.5 \textrm{M}_{\odot}$.
Detailed distributions of initial parameters are presented in Figure \ref{initial}. In the top panel,
one can see that there is very broad range of initial eccentricities -
over 60\% of all binaries have eccentricities greater than $0.1$.
The bottom panel shows the distributions of chirp masses and coalescence times. Only 12\% of binaries will merge
within the present Hubble time.

\section{Gravitational wave spectrum of eccentric binaries}
While the spectra of circular binaries have been well-studied,
we have shown above that a large fraction of population III
binaries will be eccentric. It is therefore important to investigate the possible effects of eccentricity on the gravitational wave spectra of
compact object binaries. \citep{1963PhRv..131..435P,1964PhRv..136.1224P}

\subsection{General derivation}
The gravitational wave spectrum from a binary can be
calculated as
\begin{eqnarray}
\label{eq1}
\frac{dE_{gw}}{df_{gw}}=\frac{dE_{gw}}{dt}\left( \frac{df_{gw}}{dt}\right)^{-1}.
\end{eqnarray}
The velocity on an eccentric orbit changes over its
period, hence the instantaneous orbital frequency also varies.
Thus, the system is radiating across some range of frequencies, not only
one in particular, as in the case of a circular orbit.

In the quadrupole approximation, the orbit decays
owing to gravitational wave emission and the orbital parameters change as
\begin{eqnarray}
\frac{da}{dt} & = & - \frac{64}{5} \frac{G^3 \mu M^2}{a^3 c^5} \Psi(e),
\label{dadt0} \\
\Psi(e) & = & \frac{1+73/24 e^2 + 37/96 e^4}{(1-e^2)^{7/2}},
\label{psi}
\end{eqnarray}
while  orbital frequency is
\begin{equation}
f_{orb}={1\over 2\pi} \left({GM \over a^3} \right)^{1/2},
\label{forb}
\end{equation}
where we have denoted the total mass of the binary $M$, $\mu$ is the
reduced mass,  and $a$ is the  semi-major
axis. An eccentric binary emits
gravitational waves in a spectrum of harmonics of the
orbital frequency
\begin{eqnarray}
f^n_{gw}=nf_{orb}, \\
\frac{df_{gw}}{dt}=n\frac{df_{orb}}{dt}.
\label{fngw}
\end{eqnarray}
In the case of a circular orbit, gravitational waves are
emitted in only one mode $n=2$. Inserting Eq.
\ref{forb} and Eq. \ref{dadt0} into Eq. \ref{fngw}, we obtain
\begin{eqnarray}
\label{dfdt}
\frac{df^n_{gw}}{dt}=\frac{96}{5} \left( \frac{2 \pi}{n} \right)^{8/3} \frac{{f^n_{gw}}^{11/3} G^{5/3} M_{chirp}^{5/3}}{c^5} \Psi(e),
\end{eqnarray}
where we have introduced the chirp mass, which is a quantity that determines the amplitude
and frequency dependence of the inspiral gravitational wave signal
$M_{chirp}=\mu^{3/5} M^{2/5}$.
The power of the radiation for each harmonic was calculated
by \cite{1963PhRv..131..435P} to be
\begin{eqnarray}
\frac{dE}{dt}(n)=\frac{32}{5} \frac{G^4 \mu^2 M^3}{c^5 a^5} g(n,e),
\label{dedtn}
\end{eqnarray}
where $g(n,e)$ is
\begin{eqnarray}
\label{gne}
g(n,e) & = & \frac{n^4}{32} \left\{ [ J_{n-2}(ne)-2eJ_{n-1}(ne)+\frac{2}{n}J_n(ne)
\right.  \nonumber\\
& &   +2eJ_{n+1}(ne)-J_{n+2}(ne) ]^2 \nonumber\\
& & +(1-e^2) \left[J_{n-2}(ne)-2J_n(ne)+J_{n+2}(ne) \right]^2 \nonumber\\
& &\left.  +\frac{4}{3n^2} \left[ J_n(ne) \right]^2 \right\}
\end{eqnarray}
and $J_n$ are the Bessel functions.
This result follows from the Fourier analysis of the Kepler motion
\cite{1963PhRv..131..435P}.
We obtain the instantaneous spectrum of gravitational waves from
an eccentric binary by combining Eq. \ref{eq1}-
\ref{dedtn} and Eq. \ref{dfdt}
\begin{eqnarray}
\label{dEdf}
\frac{dE}{df^n_{gw}}=\frac{\pi}{3} \frac{1}{G} \left( \frac{4}{n^2} \right)^{1/3} \frac{\left( GM_{chirp} \right)^{5/3}}{\left( f^n_{gw} \pi \right)^{1/3}} \frac{g(n,e)}{\Psi(e)}.
\end{eqnarray}
For a circular orbit this reduces to the well-known result, e.g. \citep{2001astro.ph..8028P}
\begin{eqnarray}
\label{dEdf0}
\frac{dE}{df_{gw}} \Bigg\vert_{e=0}=\frac{\pi}{3} \frac{1}{G} \frac{\left( GM_{chirp} \right)^{5/3}}{\left( f_{gw} \pi \right)^{1/3}},
\end{eqnarray}
since  $\Psi(e=0)=1$, $g(n,e=0)=1$  when $n=2$ and $g(n\neq 2,e=0)=0$.
The result has to be averaged over the lifetime of the binary. In the following
section, we discuss  two cases: the long-lived and the short-lived
binaries, as outlined below.

\subsection{Long-lived systems}
By the term long-living, we mean that the time to coalescence is longer than
the present Hubble time, which we assume for simplicity to be $t_{H}=10^4 Myr$.
We are interested in objects that have not merged.
These objects have very wide orbits and the semi-major axis of
the binary, as well as its eccentricity,
can be assumed to be constant during the entire evolution time.
The spectrum is then discrete as the
binary emits the gravitational radiation only at the frequencies specified by the
harmonics
\begin{equation}
\frac{dE}{df}=\sum_{n=2}^\infty \delta(f-n f_{orb}) \frac{dE}{d f^n_{gw}},
\end{equation}
where the spectrum in each harmonic is given
by Eq. \ref{dEdf}.

\begin{figure}
\includegraphics[scale=0.5, angle=270]{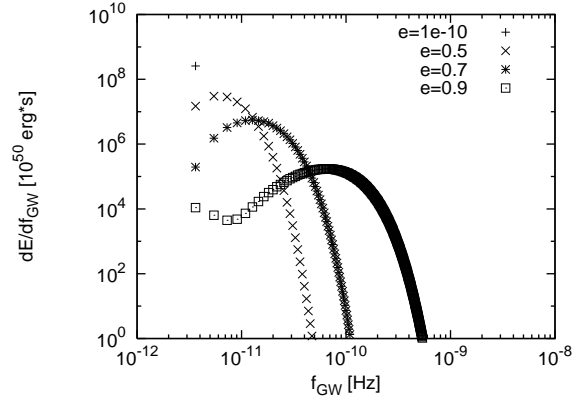}
\caption{Spectra of the GW radiation from a long-living binary ($M_1=M_2=300\, M_{\odot}$, $a=1.2\times 10^6\, R_{\odot}$) for different values of eccentricity.
A single 'plus' mark corresponds to a circular orbit, by crosses we show the case for eccentricity 0.5, stars corresponds to eccentricity 0.7, and squares to initial eccentricity
0.9.}
\label{long}
\end{figure}

To illustrate this case, we chose a binary consisting of two
massive black holes of masses
$M_1=M_2=300\, M_{\odot}$ on a wide orbit $a=1.2\times 10^6\, R_{\odot}$.
We present the spectra for a few values of the eccentricity in  Figure \ref{long}.

When the eccentricity is almost zero, we have just one point,which corresponds to a very wide, circular orbit.
The velocity is then constant and we observe only one frequency. As the eccentricity increases, we observe that
spectrum becomes wider. In addition, the maximum of the energy distribution is shifted to higher frequencies.
For the highest considered eccentricity (0.9),
there is a minimum around a frequency $10^{-11}$, which is caused by the specific structure of the function described by Eq. \ref{gne}.

\subsection{Short-lived systems}
In this group, we consider systems that can coalescence during their evolution.
The orbit changes and we observe radiation across a very wide range of frequencies.
The binary emits a continous spectrum
over its lifetime. The orbital frequency changes from $f_{ini}$, which is 
the initial orbital frequency determined by the orbital parameters
after formation of the second compact object in the binary, to
the final orbital frequency just before merger. For the final
orbital frequency, we chose the frequency for the marginally stable orbit
\begin{equation}
f_{ISCO} = \frac{c^3}{6 \sqrt{6}\pi G (M_1+M_2)}.
\end{equation}
As the binary evolves, its eccentricity decreases and its
evolution of eccentricity can be found by solving the
equation
\begin{equation}
\frac{de }{df_{orb}} = -\frac{19}{18}\frac{1}{f_{orb}} \frac{e}{(1-e^2)^{5/2}}
\left( 1+ \frac{121}{304} e^2\right).
\end{equation}
Denote the dependence of the eccentricity on the orbital frequency as $e(f_{orb})$.

The gravitational wave spectrum is
\begin{equation}
\frac{dE}{df}=\sum_{n=2}^\infty \frac{dE}{d f^n_{gw}} (e=e(f^n_{gw}/n)),
\end{equation}
where the functions $\frac{dE}{d f^n_{gw}}$ are evaluated
only for the frequencies in the range $n f_{ini}< f^n_{gw} < n f_{ISCO} $.

\begin{figure}
\includegraphics[scale=0.5, angle=270]{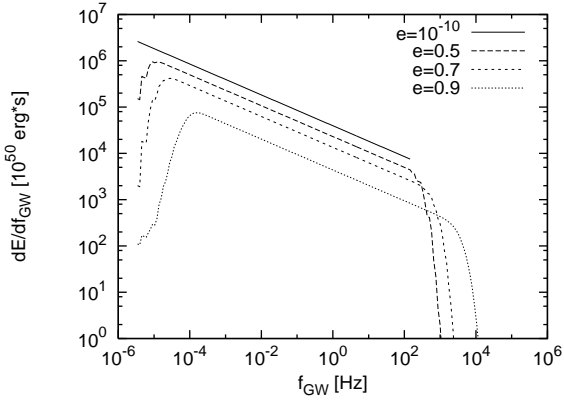}
\caption{Spectrum of the GW radiation from  short-lived binaries for different values of the initial eccentricity. Solid line corresponds to circular orbit, by dashed line we show case for eccentricity 0.5, dotted line corresponds to eccentricity 0.7 and dash-dotted line to initial eccentricity
0.9}
\label{short}
\end{figure}

We illustrate this case with a binary of the same mass as in previous section of $M_1=M_2=300 \,M_{\odot}$,
but on a much tighter orbit $a=1.2 \times 10^2\, R_{\odot}$.
We present the spectra for four different values of eccentricity in Figure \ref{short}.

For zero eccentricity, the spectrum is a simple power law.
For non-zero eccentricities, the power law spectrum acquires a tail that increases
to higher frequencies. The shape of the high frequency tail 
in the spectrum is determined by the eccentricity of the system.

\section{Gravitational wave background radiation}
We analyze the population of massive black hole binaries using the {\tt StarTrack}
 population synthesis code  \citep{2002ApJ...572..407B}.
For each system, we obtain its component masses 
$M_{1,i}$,  $M_{2,i}$, and its initial orbital parameters:
the semi-major axis $a_i$, and the eccentricity $e_i$.
We denote the total mass
of the system as $M_i=M_{1,i}+M_{2,i}$ , the reduced mass
as $\mu_i=\frac{M_{1,i} M_{2,i}}{M_{1,i}+M_{2,i}}$, and the chirp mass
$M_{chirp,i}=\mu_i^{3/5} M_i^{2/5}$.

\subsection{Evolution of the orbit}
The initial value of the semi-major axis is given by the synthesis population code and we can use it to estimate the lowest frequency
possible for any particular binary
\begin{eqnarray}
f_{i,1}=\sqrt{\frac{GM}{\pi^2 a_i^3}}.
\end{eqnarray}
The eccentricity and semi-major axis changes during the evolution
according to the two differential equations

\begin{eqnarray}
\label{dadt}
\frac{da}{dt}=- \frac{\beta}{a^3} \Psi(e) \hspace{1.5cm}
\Psi(e)=\frac{1+73/24 e^2 + 37/96 e^4}{(1-e^2)^{7/2}},
\end{eqnarray}

\begin{eqnarray}
\label{dedt}
\frac{de}{dt}=- \frac{19}{12}\frac{\beta}{a^4} \Phi(e) \hspace{1.5cm}
\Phi(e)=\frac{(1+121/304 e^2)e}{(1-e^2)^{5/2}}.
\end{eqnarray}
We have to calculate the size of the orbit after a Hubble time. For short-living systems, we
assume that the final great semi-major axis is $a_f=10^{-3} R_{\odot}$, beyond which we should consider tidal effects.
The angular momentum timescale is comparable to age of the binary, hence the system will merge eventually.
For long-living systems, we have to solve Eq. (\ref{dadt}) and
Eq. (\ref{dedt}). We can then calculate the frequency corresponding to the final size
of the orbit in the same way as we did it for initial parameters.

\begin{figure*}
\centerline{\includegraphics[scale=0.75,angle=270]{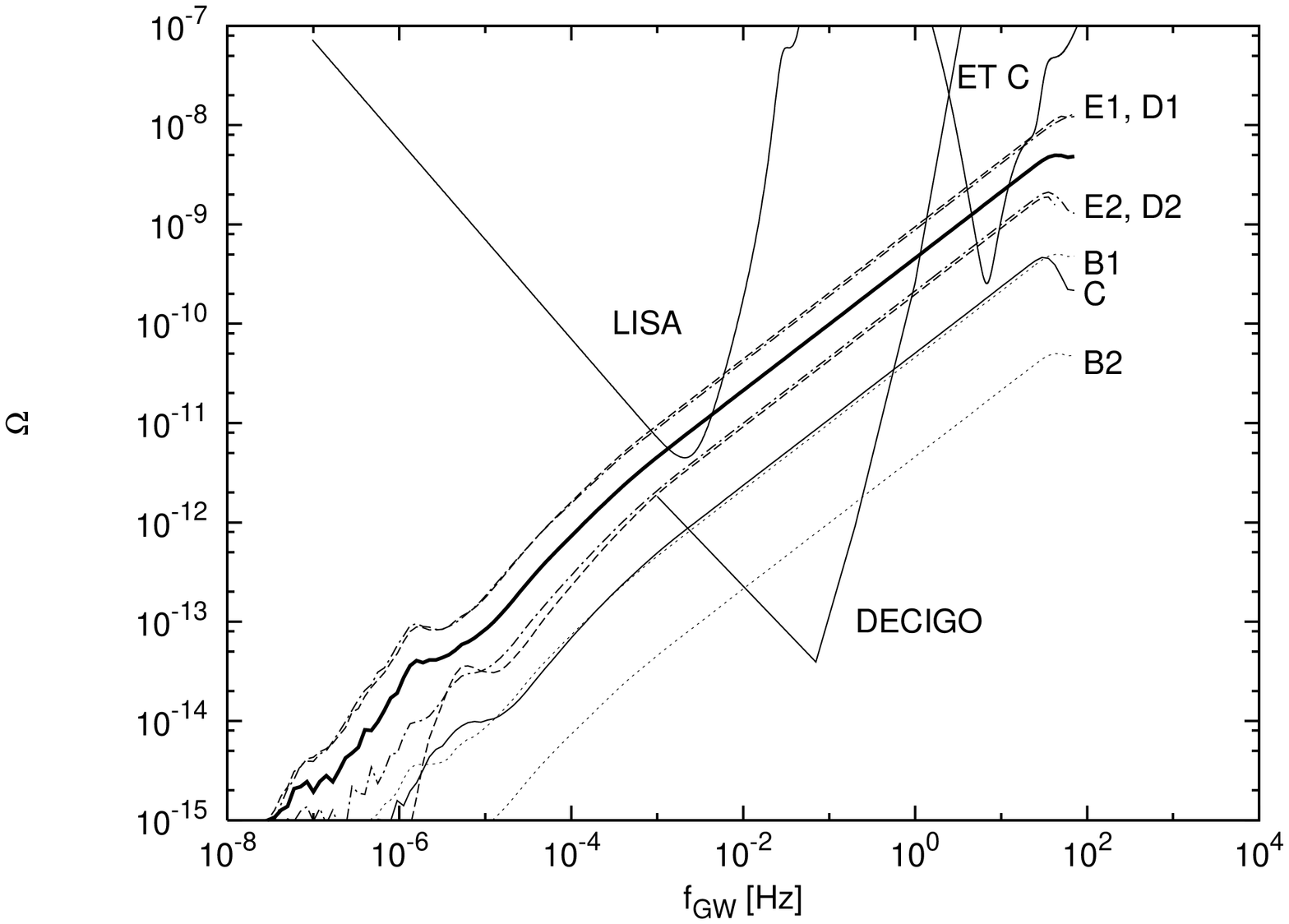}}
\centerline{\includegraphics[scale=0.75,angle=270]{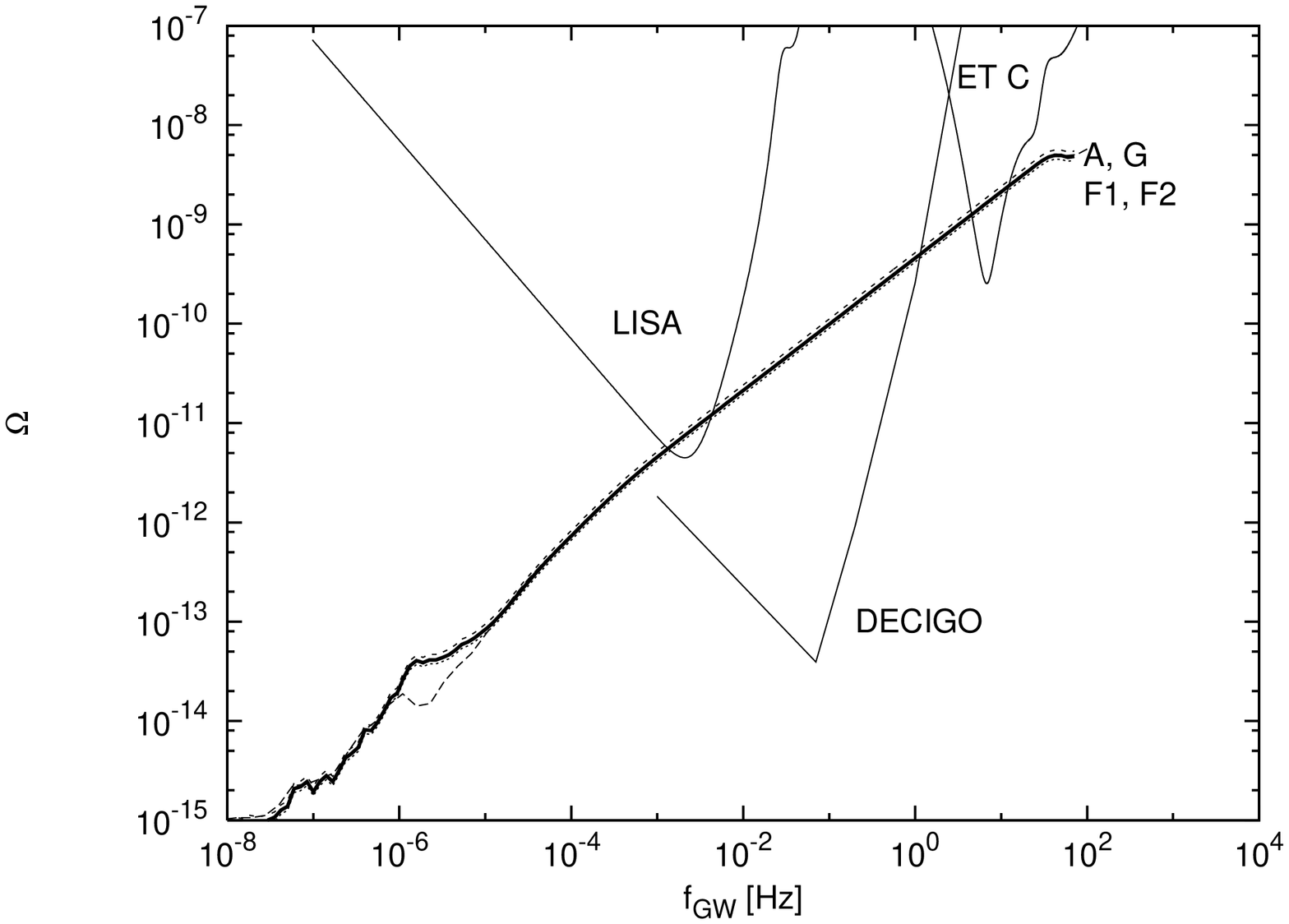}}
\caption{The gravitational wave background  from population III 
stars. The standard model results are presented as the thick solid line and denoted as A.
The thin lines are labeled with a letter denoting the model; for a description, 
see Table~\ref{table:models}. We also present the sensitivities of future gravitational wave experiments.
Bottom panel was created for clarity, because the models G, F1, and F2 are no different from the standard model.
}
\label{fig:results}
\end{figure*}

\subsection{Spectrum}
A stochastic background is described in terms of the present-day gravitational wave energy density ($\rho_{gw}$) per
logarithmic frequency interval normalized to the critical rest-mass energy density ($\rho_c c^2$) \cite{2001astro.ph..8028P}
\begin{eqnarray}
\Omega_{gw}(f)=\frac{1}{\rho_c c^2}\frac{d \rho_{gw}(f)}{d \ln(f)}.
\end{eqnarray}
The total present-day energy density contained in the gravitational radiation is related to $\Omega(f)$ and
the amplitude of the gravitational wave spectrum over a logarithmic frequency interval ($h_c(f)$) through
\begin{eqnarray}
\label{energy density 1}
\mathcal{E}_{gw}= \int_0^{\infty} \rho_c c^2 \Omega_{gw}(f) \frac{df}{f}=\int_0^{\infty} \frac{\pi}{4}
\frac{c^2}{G} f^2 h_c^2(f) \frac{df}{f}.
\end{eqnarray}
On the other hand, $\mathcal{E}_{gw}$ should be the sum of the energy radiated by all sources at all redshifts
\begin{eqnarray}
\label{energy density 2}
\mathcal{E}_{gw}= \int_0^{\infty} \int_0^{\infty} N(z) \frac{1}{1+z}f_r \frac{dE_{gw}}{df_r}dz\frac{df}{f}.
\end{eqnarray}
Since the Universe is expanding, the observed frequency is different from the emitted one and equal to $f_r=f(1+z)$, while 
$N(z)$ is the number of sources in the interval $(z,z+1)$.

Combining (\ref{energy density 1}) and (\ref{energy density 2}), we obtain
\begin{eqnarray}
\label{energy density 3}
\rho_c c^2 \Omega_{gw}(f)&=&\frac{\pi}{4} \frac{c^2}{G} f^2 h_c^2(f)= \\
&=&\int_0^{\infty} N(z) \frac{1}{1+z} \left (f_r \frac{dE_{gw}}{df_r} \right ) \Bigg\vert_{f_r=f(1+z)} dz. \nonumber
\end{eqnarray}
Using Eq. (\ref{dEdf}), we obtain equations describing the spectrum from a single harmonic for one type of source
\begin{eqnarray}
\Omega(f_{gw}^{n})&=&\frac{G^{2/3} M_{chirp}^{5/3} 2^{2/3} \pi^{2/3}}{3 n^{2/3} \rho_c c^2} \times \\ & & \times(f_{gw}^n)^{2/3} \frac{g(n,e)}{\Psi(e)} N_0 \left<(1+z)^{-1/3} \right>,
\nonumber
\end{eqnarray}
where $N_0$ is the density of one type of sources
\begin{eqnarray}
N_0=\frac{1}{N_{tot}}f_{bin}n_{pop3}, \qquad f_{bin}=\frac{2}{1+\frac{1}{f_b}}.
\end{eqnarray}
The term in the brackets is
\begin{eqnarray}
\left<(1+z)^{-1/3}\right>=\frac{1}{N_0} \int_{z_{min}}^{z_{max}} \frac{N(z)}{(1+z)^{1/3}} dz.
\end{eqnarray}
In the case of population III binaries, the 
right-hand side integral is trivial, since  we assumed that all population III stars were formed at single redshift $z=15$. Taking into account
the very short evolution timescales of massive stars population III stars,
we were able to assume that the binary compact objects were formed at the same redshift.

To obtain the energy density from all simulated objects, 
we sum over harmonics (index $n$) and over the total number of binaries (index $i$)
\begin{eqnarray}
\Omega(f)&=&\sum_{i=1}^{N} \sum_{n=2}^{\infty} \delta(f-f_{i}^{n}) \frac{G^{2/3} M_{chirp,i}^{5/3} 2^{2/3} \pi^{2/3}}{3 n^{2/3} \rho_c c^2} \times \\
& & \times (f_i^n)^{2/3} \frac{g(n,e)}{\Psi(e)} N_0 \left<(1+z)^{-1/3} \right>.
\nonumber
\label{omega}
\end{eqnarray}
The  amplitude of the gravitational wave spectrum
($h_c(f)$) and the spectral density of the gravitational wave background ($S(f)$)
are given by
\begin{eqnarray}
h_c^2(f)=\frac{4 \rho_c G}{\pi f^2} \Omega(f) &~~~& 
S(f)=\frac{h_c}{f}.
\end{eqnarray}

\section{Results}
We present the resulting gravitational wave background spectrum 
 in Figure~\ref{fig:results}. The standard model 
spectrum is shown as the thick continuous line. 
The shape of the background spectrum is determined by three 
different factors. In the low frequency regime, below  $10^{-5}$Hz ,
the background originates in long-lived systems, thus the spectrum is 
basically determined by the distribution of 
initial parameters of the binaries at the time of formation. 
In the intermediate region, between  $10^{-4}$Hz and $50$Hz 
the spectrum is due to the short-lived systems that merger within
a Hubble time. It has a typical slope of two-thirds, which corresponds to the orbit decay due to 
gravitational wave emission.
At the frequencies above  $50$Hz, the spectrum is determined by the higher harmonics arising due to eccentricity of the 
binaries. In this work, we neglect the 
merger and ringdown phases, which may alter the shape of the background spectrum 
in the region above $50$Hz.

\begin{figure*}
\centerline{\includegraphics[scale=0.9,angle=270]{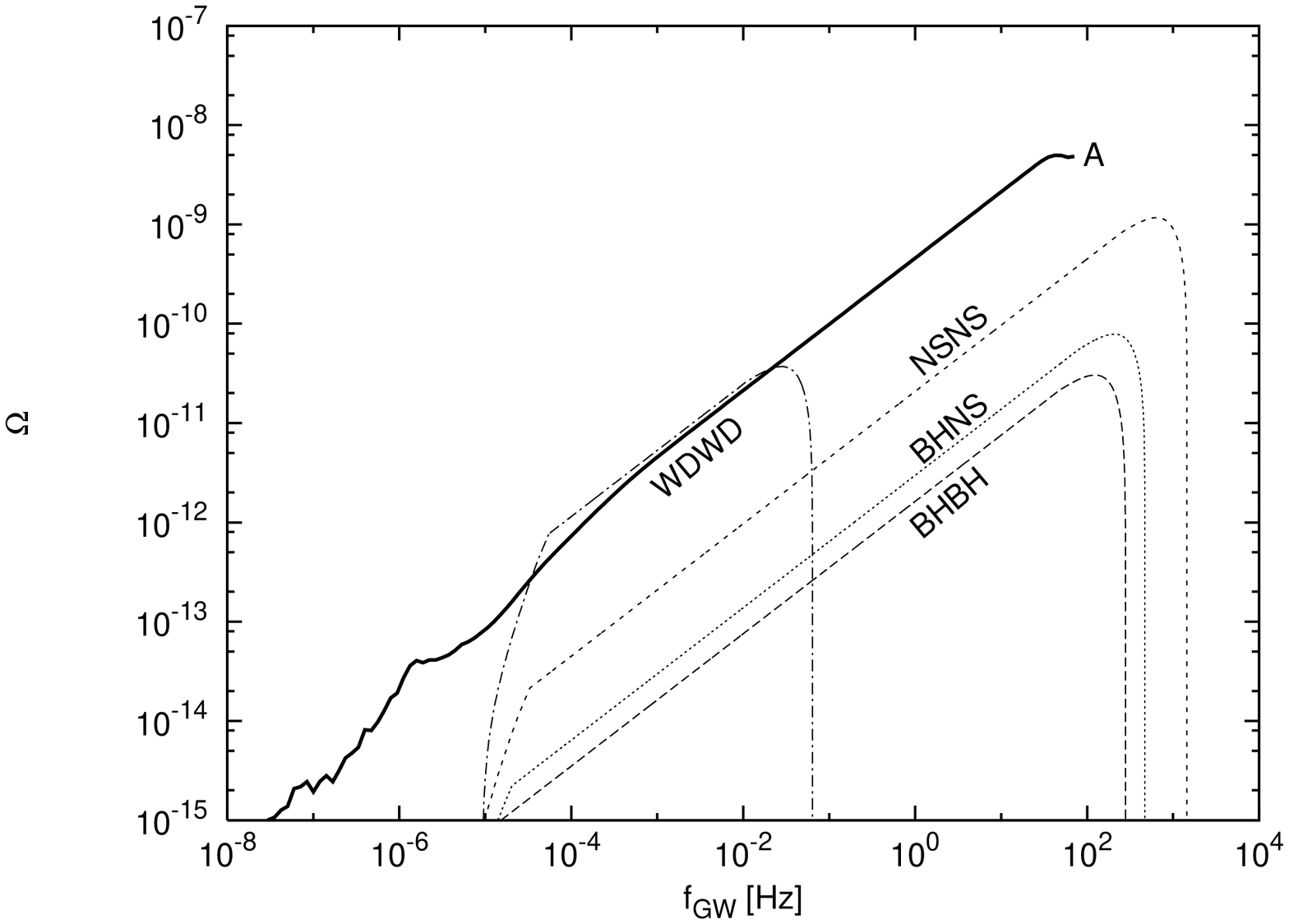}}
\caption{The gravitational wave background  from population III 
stars. Comparison with the results of \cite{2011PhRvD..84h4004R}. Solid line corresponds to our standard model A. Dotted and dashed lines
indicate the background from the different classes of compact objects calculated by \cite{2011PhRvD..84h4004R}.}
\label{fig:compar}
\end{figure*}

\subsection{Dependence on the models}
The calculation of the background involved assuming 
the values of several parameters. In this section, we present the dependence of the final result on the choice of these parameters.
The list of models is shown in Table~\ref{table:models}.
In the models B1, B2, C, D1, and D2, we vary the initial mass distributions 
of the binaries. In model D1, we increased the minimum mass
of a population III star, while in model D2 we decreased the upper end of the IMF.
Models B1 and B2 have different values of the initial 
mass function slope, and in model C we draw the two masses in the binary from the same 
IMF. Models E1 and E2 are characterized by a lower value of the 
binary fraction of $0.01$ for the first one and $0.001$ for the latter one.
In models F1 and F2, we place the binaries at different formation redshifts of
$z_{form}=10$ and $20$.  Finally, model G corresponds to the population 
of circular binaries.

\begin{table}
\caption{List of models used in the parameter study}
\begin{tabular}{rl}
Model  & Description \\ \hline \hline
A & Standard\\
B1 & Binary fraction decreased to $10^{-2}$\\
B2 & Binary fraction decreased to $10^{-3}$\\
C & Star masses drawn independently from same distribution\\
D1 & The minimum mass increased to $50M_\odot$\\
D2 & The maximum mass deceased to $100M_\odot$\\
E1 & The IMF exponent decreased to $1.5$ \\
E2 & The IMF exponent increased to $3$ \\
F1 & Formation redshift $z_{form}=10$\\
F2 & Formation redshift $z_{form}=20$\\
G & All binaries are initially circular
\end{tabular}
\label{table:models}
\end{table}

The shape of the gravitational wave background calculated with these
different assumptions is shown in  Figure \ref{fig:results}.
The effects of decreasing the binary fraction affects only 
the normalization of the spectrum without altering its shape.
Varying the initial mass function can change both the normalization 
of the spectrum, as well as the shape of the high frequency region.
Model D2 has a cutoff at frequency $10^{-6}$ Hz because of the lack of
high mass binaries.
Drawing the two stars from the same distribution leads to the formation 
of a large number of small mass binaries and the overall level of
the spectrum decreases. This is similar to increasing the IMF exponent
(see model E2). When the IMF exponent is increased (model E1),
the level of the spectrum goes up as there are more high mass binaries.
The gravitational wave spectra depend very weakly on the 
redshift of the formation of population III stars, as can clearly be seen from
Eq. \ref{omega}.
Finally, the effects of eccentricity are demonstrated by model G,
where all binaries follow initially circular orbits. This affects only the 
high frequency tail of the spectrum.

We also compare the gravitational wave background
of population III binaries with the gravitational wave backgrounds
of other types of compact binaries. \cite{2011PhRvD..84h4004R} performed such 
a comprehensive calculation. We present the standard model 
results along with the results of the \cite{2011PhRvD..84h4004R} calculation in Figure~\ref{fig:compar}.
The lower mass compact object binaries dominate the region above 
100Hz. In the region below 100Hz, the spectrum of the population III
background is an order of magnitude higher than the NSNS background 
calculated by \cite{2011PhRvD..84h4004R}.

From Figure \ref{fig:results}, one can see that the 
gravitational wave background should be clearly detectable with the 
next generation instruments. The ET will detect it, if the 
binary fraction was larger than $10^{-2}$. In the case of LISA, the 
background will show up just above the threshold of the 
band around $10^{-3}$Hz. However, for future experiments such as DECIGO,
the gravitational wave background from population III
stars could be the dominating noise source, even if 
the binary fraction was as low as $10^{-3}$. However, at this  level 
the gravitational wave background from population III stars will 
be well-buried under the background coming from 
double neutron star binaries formed at later epochs.

\section{Summary}
We have analyzed the population III 
metal-free binaries. The evolution of these binaries leads to the formation of 
binary black holes, which in turn will
be a source of gravitational waves. We have calculated the 
stochastic background from these binaries and shown 
that it can potentially be a  significant contribution to the
overall gravitational wave background. The spectrum has a 
characteristic slope of $\frac{2}{3}$ below $\approx 50$Hz, and declines 
above that frequency.

We have investigated the dependence of the shape of the gravitational 
wave spectrum on the evolutionary parameters of the 
population III binaries. The evolutionary parameters mainly affect 
the level of the spectrum, while its shape is rather insensitive to 
the parametrization. Only in the region above $50$Hz, where 
the spectrum fall off rapidly, does the shape of the spectrum 
vary with the model. In this spectral range, the 
shape is mainly affected by the initial mass function of the 
population III binaries.
For a binary fraction greater than $10^{-2}$, the 
gravitational wave background below $100$Hz is dominated 
by the contribution of population III BBH binaries.

The stochastic background from population III BBH binaries
should be detectable by the next generation instruments 
such as LISA, DECIGO, and ET.
However, the distinctive feature of this background is the 
break in the region of $50-100$Hz, which is unfortunately below the 
sensitivity of ET. Thus, it will be very difficult to distinguish 
the origin of the background if detected by any of the above-mentioned instruments.
A detailed investigation of the spectrum of the gravitational wave background 
above $50$Hz would be most interesting. It could potentially reveal
the nature of the sources of gravitational wave background, their 
mass spectrum, as well as show some effects connected with eccentricity 
of the binaries.
On the other hand, lack of detection of the stochastic 
background will lead to some constraints on the 
binary fraction and masses of population III binaries.

\acknowledgement{The authors would like to thank Tania Regimbau for a careful reading
the manuscript and helpful discussions and Pablo Rosado for providing data from his latest research.
This work was supported by the FOCUS 4/2007 Program of Foundation for Polish Science,
the Polish grants N~N203~302835,N~N203~511238, N~N203~404939, DPN/N176/VIRGO/2009, and
the Associated European Laboratory ``Astrophysics Poland-France''}

\bibliography{ecc}
\end{document}